# Tunable Correlated Chern Insulator and Ferromagnetism in Trilayer Graphene/Boron Nitride Moiré Superlattice

Guorui Chen[1,2], Aaron L. Sharpe[3,4], Eli J. Fox[4,5], Ya-Hui Zhang[6], Shaoxin Wang[2], Lili Jiang[2], Bosai Lyu[7,8], Hongyuan Li[2], Kenji Watanabe[9], Takashi Taniguchi[9], Zhiwen Shi[7,8], T. Senthil[6], David Goldhaber-Gordon[4,5]*, Yuanbo Zhang[8,10,11]*, Feng Wang[1,2,12]*

[1]Materials Science Division, Lawrence Berkeley National Laboratory, Berkeley, CA, USA.
[2]Department of Physics, University of California at Berkeley, Berkeley, CA, USA.
[3]Department of Applied Physics, Stanford University, Stanford, CA 94305, USA.
[4]Stanford Institute for Materials and Energy Sciences, SLAC National Accelerator Laboratory, 2575 Sand Hill Road, Menlo Park, California 94025, USA.
[5]Department of Physics, Stanford University, Stanford, California 94305, USA.
[6]Department of Physics, Massachusetts Institute of Technology, Cambridge, Massachusetts, USA.
[7]Key Laboratory of Artificial Structures and Quantum Control (Ministry of Education), School of Physics and Astronomy, Shanghai Jiao Tong University, Shanghai, China.
[8]Collaborative Innovation Center of Advanced Microstructures, Nanjing, China.
[9]National Institute for Materials Science, 1-1 Namiki, Tsukuba, 305-0044, Japan.
[10]State Key Laboratory of Surface Physics and Department of Physics, Fudan University, Shanghai 200433, China.
[11]Institute for Nanoelectronic Devices and Quantum Computing, Fudan University, Shanghai 200433, China.
[12]Kavli Energy NanoSciences Institute at the University of California, Berkeley and the Lawrence Berkeley National Laboratory, Berkeley, CA, USA.

*Correspondence to: fengwang76@berkeley.edu, zhyb@fudan.edu.cn, goldhaber-gordon@stanford.edu

**Studies on two-dimensional electron systems in a strong magnetic field first revealed the quantum Hall (QH) effect[1], a topological state of matter featuring a finite Chern number (*C*) and chiral edge states.[2,3] Haldane later theorized that Chern insulators with integer QH effects could appear in lattice models with complex hopping parameters even at zero magnetic field.[4] The ABC-trilayer graphene/hexagonal boron nitride (TLG/hBN) moiré superlattice provides an attractive platform to explore Chern insulators because it features nearly flat moiré minibands with a valley-dependent electrically tunable Chern number.[5,6] Here we report the experimental observation of a correlated Chern insulator in a TLG/hBN moiré superlattice. We show that reversing the direction of the applied vertical electric field switches TLG/hBN's moiré minibands between zero and finite Chern numbers, as revealed by dramatic changes in magneto-transport behavior. For topological hole minibands tuned to have a finite Chern number, we focus on 1/4 filling, corresponding to one hole per moiré unit cell. The Hall resistance is well quantized at $h/2e^2$, i.e. *C* = 2, for |*B*| > 0.4 T. The correlated Chern insulator is ferromagnetic, exhibiting significant magnetic hysteresis and a large anomalous Hall signal at zero magnetic field. Our discovery of a *C* = 2 Chern insulator at zero magnetic field should open up exciting opportunities for discovering novel correlated topological states, possibly with novel topological excitations,[7] in nearly flat and topologically nontrivial moiré minibands.**

Moiré superlattices in van der Waals heterostructures have emerged as a powerful tool for engineering novel quantum phenomena, where the periodic moiré potential defines new length and energy scales.[8–10] Notably, nearly flat electronic bands can be realized in different moiré superlattice systems, which offer exciting opportunities to realize a wide variety of correlation physics.[5,6] For example, correlated insulators and superconductivity have been reported in magic-angle twisted bilayer graphene[11–13] and in TLG/hBN moiré superlattices[14,15], and spontaneous ferromagnetism and an anomalous Hall effect, apparently corresponding to an incipient Chern insulator, have been observed in twisted bilayer graphene with an aligned hBN layer[16]. Recent theories suggest that novel correlated topological phenomena can emerge in such graphene moiré superlattices, where a non-trivial band topology coexists with the nearly flat moiré miniband.[5,6,17] The TLG/hBN moiré superlattice provides a particularly attractive platform to explore correlated topological phenomena because not only the electron density but also the bandwidth and topology of the moiré minibands can be conveniently controlled by electrostatic gating.[5,6,14]

Here we report experimental observation of a correlated Chern insulator and ferromagnetism in TLG/hBN. Upon tuning the vertical displacement field, we show that the magneto-transport behavior in a TLG/hBN moiré superlattice exhibits distinct behaviors for trivial minibands ($C = 0$) compared to topological minibands ($C \neq 0$). A correlated Chern insulator with $C = 2$ quantum anomalous Hall (QAH) effect[18] emerges around 1/4 filling of the topological hole miniband when the bandwidth is sufficiently narrowed by applying a displacement field in one direction. The correlated Chern insulator spontaneously breaks time-reversal symmetry, exhibiting strong ferromagnetic hysteresis and a zero-field anomalous Hall (AH) resistance over 8 k$\Omega$. The experimentally observed $C = 2$ Chern band can be understood theoretically by incorporating electron-electron interaction effects on the quasi-particle band structure of TLG/hBN moiré minibands.

The TLG/hBN moiré superlattice devices are fabricated following the method described in Ref. 7. Briefly, the ABC-TLG domain is identified by scanning near-field infrared nanoscopy[19], and is isolated from adjacent ABA domains by atomic force microscope cutting[20]. The isolated ABC-TLG is then encapsulated in exfoliated hBN crystals, where one hBN crystal is aligned with the TLG to form the moiré superlattice. The TLG/hBN heterostructures are fabricated into a Hall bar geometry with one-dimensional edge contacts, a metal top gate, and a degenerately doped silicon bottom gates following standard nanofabrication procedures[21]. A schematic and optical image of the final device are shown in Figs. 1a and 1b inset. Gate voltages $V_t$ and $V_b$ are applied to the metal top gate and the Si bottom gate, respectively. The dual-gate configuration allows us to independently control the doping and the miniband bandwidth of the TLG/hBN heterostructure[22–24]: the doping relative to the charge neutrality point (CNP) is set by $n = (D_b - D_t)/e$, and the miniband bandwidth is tuned by the applied vertical displacement field $D = (D_b + D_t)/2$. Here $D_b = +\varepsilon_b(V_b - V_b^0)/d_b$ and $D_t = -\varepsilon_t(V_t - V_t^0)/d_t$ are the vertical displacement field below and above the TLG/hBN moiré superlattice, respectively, $\varepsilon_{b(t)}$ and $d_{b(t)}$ are the dielectric constant and thickness of the bottom (top) dielectric layers, and $V_{b(t)}^0$ is the effective offset in the bottom (top) gate voltages caused by environment-induced carrier doping. The longitudinal resistivity $\rho_{xx}$ is obtained by $\rho_{xx} = (W/L) V_{xx}/I$, where $W = 1$ μm is the channel width and $L = 4$ μm is the channel length, and the Hall resistivity $\rho_{yx}$ is obtained by $\rho_{yx} = V_{yx}/I$ (the measurement configuration for the longitudinal and Hall voltages, $V_{xx}$ and $V_{yx}$, respectively, is shown in Fig. 1a).

As the voltages applied to the gates are tuned, measurements of $\rho_{xx}$ reveal several resistance peaks (Fig. 1b) in the TLG/hBN device across the parameter space controlled by $V_t$ and $V_b$. In addition to the peaks corresponding to band insulating states at the CNP and fully filled point (FFP), tunable correlated insulator states emerge at 1/4 filling and 1/2 filling of the hole minibands (i.e. one and two holes per moiré unit cell) when a finite displacement field |D| narrows the moiré minibands. There are two apparent asymmetries of the correlated insulator states in TLG/hBN at 1/4 and 1/2 charge fillings: between the electron and hole minibands, and between positive and negative D.

Prominent correlated insulator states are observed in the hole minibands but not in the electron minibands because the hole miniband has a much smaller bandwidth for finite $|D|$.[14] The asymmetry between the positive and negative $D$ fields arises from the fact that the moiré superlattice exists only between the TLG and the bottom hBN in this device (Fig. 1a). Interestingly, the direction of the displacement field has been predicted to determine not only the relative bandwidth but also the topology of the hole miniband in such devices: a positive $D$ leads to a trivial hole miniband with $C = 0$ and smaller bandwidth, while a negative $D$ leads to a topological hole miniband with $C \neq 0$ and larger bandwidth.[5,6] The difference in bandwidth and topology is reflected in the larger resistivity peaks of the correlated insulators in the trivial hole miniband with positive $D$, because holes are easier to localize in the narrower trivial band than in the broader topological band with negative $D$.[17]

To better probe the topological aspects of the moiré minibands, we turn to magneto-transport studies. Figs. 2a and 2b display the color plot of $\rho_{xx}$ as a function of the hole doping and the vertical magnetic field, $B$, for $D = -0.5$ V/nm (at $T = 0.06$ K) and $D = 0.4$ V/nm (at $T = 1.5$ K), respectively. Figs. 2c and 2d show the corresponding Hall resistivity $\rho_{yx}$ data. Experimental data for $D = -0.5$ V/nm at 1.5 K exhibit qualitatively similar behavior to those at 0.06 K (See Extended Data Fig. 1). We have used $n_0 = 5.25 \times 10^{11}$ cm$^{-2}$ as the unit of carrier density, which corresponds to one hole per moiré lattice site (i.e. 1/4 filling). The magneto-transport data exhibit distinct behaviors for the topological moiré miniband at negative $D$ and the trivial miniband at positive $D$ [5,6]. Specifically, a strong QH state emerges from the 1/4 filling point for $D = -0.5$ V/nm but not for $D = 0.4$ V/nm. The dashed line in Fig. 2a traces the minimum in $\rho_{xx}$ following the relation $n = \nu eB/h$ for $\nu = 2$. This QH state is well developed at very low magnetic field, and originates from the 1/4 filling resistive state at zero magnetic field (Fig. 2a). At the same time, $\rho_{yx}$ is very large at weak magnetic fields and exhibits a jump in value when the magnetic field switches sign across $B = 0$ T (Fig. 2c). In contrast, stronger correlated insulator states are observed for $D = 0.4$ V/nm, but no signatures of quantum oscillations or QH effects are present (Fig. 2b). In addition, Fig. 2d shows that the Hall resistivity signal tends to be rather small for all hole doping at $D$ = 0.4 V/nm. (The relatively large $\rho_{yx}$ signals at 1/4 and 1/2 fillings are artifacts due to crosstalk from the large $\rho_{xx}$ of the correlated insulator states, and they do not change sign when the magnetic field is reversed.)

Fig. 2e and 2f show $\rho_{yx}$ and $\rho_{xx}$ as a function of density for a few representative magnetic field values, corresponding to horizontal line cuts in Figs. 2c and 2a, respectively. $\rho_{yx}$ is well quantized for magnetic field larger than 0.4 T at the value of 13.0 ± 0.2 kΩ, i.e. the expected quantized value of $h/2e^2$ = 12.9 kΩ is within the empirical uncertainty. $\rho_{xx}$ exhibits a corresponding minimum in the QH state, with a minimum resistivity less than 60 Ω at 2 T. Fig. 2g further displays $\rho_{yx}$ and $\rho_{xx}$ as a function of the magnetic field along the QH state following the dashed line in Fig. 2c, with the inset showing a zoomed-in plot of $\rho_{yx}$ between 0 to 0.2 T. $\rho_{yx}$ smoothly reaches the quantized value at 0.4 T. $\rho_{yx}$ maintains a large though not quantized value all the way to zero magnetic field, and a large jump of $\rho_{yx}$ is observed when the magnetic field changes sign.

The $\nu$ = 2 QH state at 1/4 filling at $D$ = -0.5 V/nm cannot be explained by a conventional integer QH effect from single-particle Landau levels. Instead, we argue it represents a QAH state from a correlated Chern insulator. Firstly, this QH state only exists at negative $D$, where the miniband is predicted to have a non-trivial Chern number, and is absent at the positive $D$, where the band is predicted to be trivial. Secondly, it is well established that the lowest single-particle Landau level in ABC-TLG should be a $\nu$ = 3 state due to a winding number of 3 close to the valence band maximum.[25] Thirdly, only one QH state is observed anywhere, and the quantized Hall resistivity appears starting at very low magnetic field. If the observed QH state is from the lowest single-particle Landau level, similar Landau levels should also exist close to the CNP and the 1/2 filling correlated insulators, and higher Landau levels should be observable. All our data can be naturally explained by a $\nu$ = 2 Chern insulator state at 1/4 filling. Such a $C$ = 2 correlated Chern insulator should feature quantized Hall resistivity $\rho_{yx}$ and a corresponding magnetic field dependent carrier density based on the Streda formula[26]. This Chern insulator at 1/4 filling is a strongly correlated state that breaks the valley degeneracy and fills only the $C = 2$ electronic band in one valley. The nearly flat and tunable moiré minibands in the TLG/hBN moiré heterostructure are critical for the realization of such a correlated topological state.

The correlated Chern insulator, persisting to zero magnetic field, spontaneously breaks the time-reversal symmetry and can generate valley flavor ferromagnetism at 1/4 filling. Indeed, ferromagnetism and strong AH signals emerge from the Chern insulator state at zero magnetic field. Fig. 3a shows the temperature dependent Hall resistivity when a small perpendicular $B$ is swept between -0.1 to 0.1 T. The Hall resistivity displays a clear AH signal with strong ferromagnetic hysteresis. At $B = 0$ T, $\rho_{yx}$ is nonzero and depends on the magnetic field sweep direction, a defining ferromagnetic feature. At the base temperature of $T = 0.06$ K, the AH signal reaches a maximum of $\rho_{yx}^{AH} = 8$ k$\Omega$ and a coercive field as large as $B_c = 30$ mT. The inset in Fig. 3a shows the temperature dependence of $\rho_{yx}^{AH}$ and $B_c$: both signals decrease monotonically with increasing temperature, reaching zero at $T = 3.5$ K. The zero magnetic field $\rho_{yx}^{AH}$ is already close to 12.9 k$\Omega$. An almost perfect quantization of the $C = 2$ QAH[18] Chern insulator appears at a magnetic field as low as 0.4 T.

The ferromagnetism is tunable by $n$ and $D$ and only appears in a limited parameter space of $n$ and $D$. In Fig. 2e, $\rho_{yx}$ near $n = n_0$ presents different signs at $B = -4$ mT and 6 mT, which is much smaller than $B_c$. For this measurement, the magnetic field is fixed and carrier density is swept from a non-ferromagnetic state to a ferromagnetic state, which leads to $\rho_{yx}$ with different signs even in small positive and negative magnetic fields. Clearer $n$-dependence of $\rho_{xx}$, $B_c$ and $\rho_{yx}^{AH}$ at $D = -0.5$ V/nm at the base temperature are shown in Fig. 3b by sweeping magnetic field at different fixed $n$. $B_c$ and $\rho_{yx}^{AH}$ both have maximum values close to $n = n_0$. However, $\rho_{yx}^{AH}$ shows a stronger carrier doping dependence and decreases to almost zero at $n = n_0 \pm 0.2n_0$, while $B_c$ decreases to zero at $n = n_0 \pm 0.35\ n_0$. $\rho_{xx}$ displays a rather unusual behavior with both a resistance peak and a resistance dip close to $n = n_0$, the origin of which requires further experimental and theoretical study. Fig. 3c shows the $D$-dependence of $\rho_{xx}$, $B_c$ and $\rho_{yx}^{AH}$ at $n = n_0$. $\rho_{xx}$ shows a maximum at $D = -0.5$ V/nm, which might be due to the narrowest bandwidth and strongest correlation effects at this displacement field.[14] $\rho_{yx}^{AH}$ also shows a maximum at $D = -0.5$ V/nm, suggesting the importance of electron-electron correlation to the observed AH signal. A finite $\rho_{yx}^{AH}$ can be observed with $D$ between -0.3 V/nm and -0.57 V/nm. A non-zero $B_c$ is present in the same $D$ range, although the maximum of $B_c$ appears at $D = -0.45$ V/nm.

The observed $C = 2$ correlated Chern insulator can be understood theoretically from the topological moiré minibands when the electron-electron interactions are considered. Previous theoretical calculations predict a valley Chern number $C = 3$ for the single-particle hole miniband for negative $D$,[5,6] but our results suggest that interaction effects can renormalize the valley Chern number. Figs. 4a and 4b show the single-particle band structures of the lowest few moiré minibands in TLG/hBN moiré superlattices for positive and negative displacement fields. For the negative $D$ (supporting a non-zero valley Chern number), the valence band overlaps with the remote lower band (see Fig. 4a). We incorporate the interaction effects in a Hartree Fock theory. When the valence band is close to the band below (at large $|D|$) or when the interaction strength is sufficiently strong (with small dielectric constant), the self-energy corrections mix the valence band and the lower band, leading to reduction of the Chern number to $C = 2$. As shown in Fig.4c, when the dielectric constant is around 4 (effective screening from the dielectric constant of hBN), the valley Chern number is expected to be 2 for a large range of displacement field values.

Next, we discuss the nature of the correlated insulator at 1/4 filling for the topological side. Because of the Wannier obstruction caused by the valley Chern number, a standard Mott insulator with localized charge is impossible.[17] Therefore, the physics on the topological side is essentially different from the trivial side despite the similarity between the band structures. Because a narrow Chern band is analogous to a Landau level, the physics on the topological side is similar to that of quantum Hall systems with spin and valley degeneracies. At 1/4 filling, when the band is sufficiently flat, a single fully-filled spin and valley polarized Chern band is favored[6], similar to "quantum Hall ferromagnetism" found in Landau levels[27]. A valley polarized Chern insulator matches the current transport experiment quite well. Ideally, a fully filled Chern band leads to quantized Hall conductivity $\sigma_{yx} = 2e^2/h$. At zero magnetic field, domains formed by the two degenerate valleys can cause a finite $\rho_{xx}$ on order of $h/e^2$. Upon increasing the magnetic field to ~0.4 T, the valley Zeeman coupling[17] can align the domains and lead to perfect quantization of the Hall conductivity.

Ferromagnetism and an anomalous Hall effect have also recently been observed in a near-magic-angle twisted bilayer graphene (TBG) moiré superlattice at 3/4 filling of the conduction miniband, where an hBN cladding layer appears to be aligned with the proximate graphene[16]. These effects are likely to have a similar origin in both systems, where electron-electron interactions create a spontaneous valley polarization in the nearly flat and topological moiré minibands.[28,29] However, the Hall resistance in the incipient Chern insulator state of the TBG device of Ref. 15 is not quantized even in finite applied magnetic field, possibly due to significant twist-angle disorder[30].

Our observation of tunable $C = 2$ Chern insulator in TLG/hBN moiré superlattice can open up exciting possibilities to explore novel correlated topological states in van der Waals moiré heterostructures. For example, fractional Chern insulators and non-Abelian states could emerge from strong correlations in nearly flat topological minibands when the quality of moiré heterostructures is further improved. In particular, the flat $C = 2$ Chern band has the potential to host completely new fractional Chern insulator states beyond the fractional quantum Hall paradigm[31,32], such as a wormhole-like "genon" state that change the geometry of space with non-Abelian exchange statistics[33].

**Acknowledgments** We thank Yujun Yu and Mengqiao Sui for measurement assistances, and acknowledge helpful discussions with M. Zaletel, E. Altman, J. Jung, and M. A. Kastner. G.C. and F.W. were supported as part of the Center for Novel Pathways to Quantum Coherence in Materials, an Energy Frontier Research Center funded by the U.S. Department of Energy, Office of Science, Basic Energy Sciences. A.S. was supported by a National Science Foundation Graduate Research Fellowship and a Ford Foundation Predoctoral Fellowship. E.F., and D.G.-G.'s work on this project were supported by the U.S. Department of Energy, Office of Science, Basic Energy

Sciences, Materials Sciences and Engineering Division, under Contract No. DE-AC02-76SF00515. Dilution fridge support: Low-temperature infrastructure and cryostat support were funded in part by the Gordon and Betty Moore Foundation through Grant No. GBMF3429. Part of the sample fabrication was conducted at Nano-fabrication Laboratory at Fudan University. Part of the measurement was performed in Oxford Instrument Nanoscience Shanghai Demo Lab. Y.Z. acknowledges financial support from National Key Research Program of China (grant nos. 2016YFA0300703, 2018YFA0305600), NSF of China (grant nos. U1732274, 11527805, 11425415 and 11421404), and Strategic Priority Research Program of Chinese Academy of Sciences (grant no. XDB30000000). Z.S. acknowledges support from National Key Research and Development Program of China (grant number 2016YFA0302001) and National Natural Science Foundation of China (grant number 11574204, 11774224). T.S was supported by NSF grant DMR-1608505, and partially through a Simons Investigator Award from the Simons Foundation. K.W. and T.T. acknowledge support from the Elemental Strategy Initiative conducted by the MEXT, Japan and the CREST (JPMJCR15F3), JST.

**Author contributions**
F.W. and G.C. conceived the project. F.W., Y.Z., D.G.-G. and T.S. supervised the project. G.C. fabricated samples and performed transport characterizations at temperature above 1 K to first identify the $C = 2$ quantum Hall states. G.C., A.S. and E.F. performed ultra-low temperature transport measurements. G.C., L.J., B.L., H.L. and Z.S. prepared trilayer graphene and performed near-field infrared and AFM measurements. K.W. and T.T. grew hBN single crystals. YH.Z. and T.S. calculated the band structures and Chern numbers. G.C., A.S., E.F., YH.Z., T.S., D.G.-G., Y.Z. and F.W. analyzed the data. G.C., YH.Z., T.S. and F.W. wrote the paper, with input from all authors.

**Author Information**
The authors declare no competing financial interests. Correspondence and requests for materials should be addressed to F.W. (fengwang76@berkeley.edu), Y.Z. (zhyb@fudan.edu.cn) and D.G.-G. (goldhaber-gordon@stanford.edu).

**Main figure legends**
**Figure 1. TLG/hBN moiré superlattice and tunable Chern bands. a**, Schematic of the dual-gated ABC-TLG/hBN moiré superlattice Hall bar device and measurement configuration. The moiré pattern exists between TLG and bottom hBN. **b**, Color plot of the longitudinal resistivity $\rho_{xx}$ as a function of $V_t$ and $V_b$ at $T = 1.5$ K. The arrows show the direction of changing doping (*n*) and displacement field (*D*), respectively. In addition to the band insulating states (characterized by the resistance peaks) at charge neutral point (CNP) and fully filled point (FFP), tunable correlated insulator states also emerge at 1/4 filling and 1/2 filling of the hole minibands at large displacement field |*D*|. It was predicted theoretically that the hole miniband is topological (i.e. Chern number $C \neq 0$) for $D < 0$ and trivial ($C = 0$) for $D > 0$.

**Figure 2. Quantum Hall effect from the correlated $C = 2$ Chern insulator. a,c** Color plot of $\rho_{xx}$ and $\rho_{yx}$ as a function of carrier density and magnetic field for the topological hole miniband at $D = -0.5$ V/nm and $T = 0.06$ K. The experimental data at $T = 1.5$ K are qualitatively similar. **b,d**, Corresponding $\rho_{xx}$ and $\rho_{yx}$ plots for the trivial hole miniband at $D = 0.4$ V/nm and $T = 1.5$ K. $n_0$ corresponds to the carrier density of 1/4 filling of first miniband. No QH signatures are present in the trivial hole miniband, while a $\nu = 2$ QH effect characterized by a minimum of $\rho_{xx}$ and a quantized $\rho_{yx}$ emerges from 1/4 filling of the topological hole miniband. **e,f**, Horizontal line cuts of **a** and **c**, respectively. **e** shows that $\rho_{yx}$ is well quantized beyond $B = 0.4$ T. An offset of 2.5 k$\Omega$ is applied for each stack in **f**. **g**, Line cut of $\rho_{xx}$ and $\rho_{yx}$ along the QH state (denoted by the dashed line in **a** and **e**) shows that $\rho_{yx}$ reaches a quantized value of $\nu = 2$ at 0.4 T, and a large $\rho_{yx}$ persists down to zero field. It represents a QAH state for the $C = 2$ correlated Chern insulator at 1/4 filling.

**Figure 3. Anomalous Hall effect and ferromagnetism. a**, Magnetic field dependent $\rho_{yx}$ at 1/4 filling and $D = -0.5$ V/nm at different temperatures. The Hall resistivity displays a clear AH signal with strong ferromagnetic hysteresis. At the base temperature of $T = 0.06$ K, the AH signal can be as high as $\rho_{yx}^{AH} = 8$ k$\Omega$ and the coercive field is $B_c = 30$ mT. Inset: Extracted coercive field $B_c$ and AH signal $\rho_{yx}^{AH}$ as a function of temperature. **b**, The evolution of $\rho_{xx}$, $B_c$ and $\rho_{yx}^{AH}$ as a function of hole doping at $D = -0.5$ V/nm, $T = 0.06$ K. The strongest AH signal is observed close to $n = n_0$. **c**, The evolution of $\rho_{xx}$, $B_c$ and $\rho_{yx}^{AH}$ as a function of the displacement field $D$ at $n = n_0$, $T = 0.06$ K. The strongest AH signal is observed when the device is most insulating (i.e. largest $\rho_{xx}$).

**Figure 4. Calculated Chern number including the electron-electron interaction effects. a,b**, Calculated single-particle band structure of the TLG/hBN moiré superlattice for $\Phi = -25$ meV and 25 meV, respectively. Here $\Phi$ is the energy difference between the top and bottom layer of TLG, and $\Phi = -25$ meV corresponds to vertical displacement field around $D = -0.5$ V/nm. The red line highlights the topological hole miniband for $\Phi = -25$ meV. **c**, Calculated Chern number of the hole miniband as a function of the energy difference $\Phi$ and the effective dielectric constant $\varepsilon_{hBN}$ after including the electron-electron interaction effects using the Hartree-Fock approximation. The resulting band Chern number can be 2 for parameters close to the experimental device where $\Phi \sim -25$ meV and $\varepsilon_{hBN} \sim 4$.

**Methods**

**Transport measurements** The ultralow temperature measurement is performed in a dilution refrigerator. Low temperature electronic filtering, including microwave filters, low-pass RC filters, and thermal meanders, is used to anchor the electron temperature as well as to prevent quasiparticle excitations from high frequency noise. Stanford Research Systems SR830 lock-in amplifiers with NF Corporation LI-75A voltage preamplifiers are used to measure the resistivity of the device with an AC bias current of 0.5 nA at a frequency of 7 Hz.

**Calculation for TLG/hBN moiré superlattice** A simple argument in Ref. 7 shows that the valley Chern number must jump by 3 when switching the direction of the displacement field if the super-lattice gap does not close during this process. A direct numerical calculation gives $C = 3$ for the $D < 0$ side. Thus, due to interactions, a symmetry breaking state with spin-valley polarization is stabilized, then a Chern insulator with $C = 3$ is expected. We first improve our theoretical modeling of the band structure by adding various remote-hopping terms at single particle level and incorporating interaction effects, as shown in Extended Data Fig. 2. As we will show, $C = 3$ is still robust in our more sophisticated model of single particle band structure. Interaction effects turn out to be necessary to explain the reduction of the Chern number.

The ABC-TLG is modeled by a six band model. We use the following parameters[34]:

$$(v, \gamma_1, \gamma_2, v_3, v_4) = (2676, 380, 8.3, 260, 104) \, meV$$

Then the Hamiltonian for the valley + is:

$$H_0(\mathbf{k}) = \begin{pmatrix} \Phi_V/2 & v(k_x - ik_y) & -v_4(k_x - ik_y) & v_3(k_x + ik_y) & 0 & \gamma_2 \\ v(k_x + ik_y) & \Phi_V/2 & \gamma_1 & 0 & 0 & 0 \\ -v_4(k_x + ik_y) & \gamma_1 & 0 & v(k_x - ik_y) & -v_4(k_x - ik_y) & v_3(k_x + ik_y) \\ v_3(k_x - ik_y) & 0 & v(k_x + ik_y) & 0 & \gamma_1 & 0 \\ 0 & 0 & -v_4(k_x + ik_y) & \gamma_1 & -\Phi_V/2 & v(k_x - ik_y) \\ \gamma_2 & 0 & v_3(k_x - ik_y) & 0 & v(k_x + ik_y) & -\Phi_V/2 \end{pmatrix}$$

The aligned hBN layer in the bottom provides a moiré hopping term which folds the original band structure to a small mini Brillouin zone. We use the same model of the moiré hopping term as in Ref. 35.

Using the above model we get the band structures shown in Fig.4 of the main text. For $\Phi_V = -25$ meV, we get a narrow valence band with Chern number $|C| = 3$. We have tried to change the various hopping parameters and the potential difference. However, the Chern number is always equal to 3 and we conclude that $C = 2$ cannot be reproduced in the single particle level.

In Fig. 4b of the main text, we can see that though at each K-point the valence band is isolated from the band below, they overlap in energy (in other words though there is a direct gap there is no indirect gap). More precisely the system is in a compensated semimetal phase at the fully filled point. This is in agreement with the experimental measurement. In the $D < 0$ side, electrons are pushed to be away from the aligned bottom hBN layer. Therefore the moiré superlattice potential has a weaker effect and the superlattice gap is small.

Given that the valence band is not isolated from the remote band below, an interaction induced self-energy can renormalize the band structure and maybe even band topology. To incorporate this effect, we perform a self-consistent Hartree-Fock calculation by only keeping the valence band and the remote band. The interacting Hamiltonian is

$$H = \sum_{k;a,\sigma} \sum_m \xi_{a,m}(\mathbf{k}) c^\dagger_{a,\sigma;m}(\mathbf{k}) c_{a,\sigma;m}(\mathbf{k})$$

$$+ \frac{1}{N} \sum_{\mathbf{q}} \sum_{\mathbf{k}_1,\mathbf{k}_2;a_1,\sigma_1,a_2,\sigma_2} c^\dagger_{a_1,\sigma_1;m_1}(\mathbf{k}_1+\mathbf{q}) c^\dagger_{a_2,\sigma_2;m_2}(\mathbf{k}_2-\mathbf{q}) c_{a_2,\sigma_2;n_2}(\mathbf{k}_2) c_{a_1,\sigma_1;n_1}(\mathbf{k}_1) V(\mathbf{q}) \lambda_{a_1;m_1 n_1}(\mathbf{k}_1,\mathbf{q}) \lambda_{a_2;m_2 n_2}(\mathbf{k}_2,-\mathbf{q})$$

where $a = \pm$ is the valley index and $m = 0,1$ is the band index labeling the valence band and the remote band. $\sigma = \uparrow, \downarrow$ is the spin index. $c_{a,\sigma;m}(\mathbf{k})$ is the creation operator corresponding to the band $m$ for the valley $a$ and spin $\sigma$. The two terms in the Hamiltonian are kinetic term and interaction term. $V(q)$ is the screened Coulomb interaction controlled by the renormalization factor of the dielectric constant. In the interaction we have included the form factors $\lambda_{a;mn}$ to incorporate the Berry curvature of the Bloch wavefunctions[6]. Then Hartree-Fock self-energy can be obtained from the self-consistent equations:

$$\sum\nolimits^H_{a,mn}(\mathbf{k}) = \frac{1}{N} \sum_{a_2,\sigma,m'} \sum_{\mathbf{G}} \sum_{\mathbf{k}_2} \langle c^\dagger_{a_2 \sigma;m'}(\mathbf{k}_2) \rangle V(G) \lambda_{a;mn}(\mathbf{k},\mathbf{G}) \lambda_{a_2;m'm'}(\mathbf{k}_2,-\mathbf{G})$$

$$\sum\nolimits^F_{a,mn}(\mathbf{k}) = -\frac{1}{N} \sum_{m'} \sum_{\mathbf{q}} \langle c^\dagger_{a\sigma;m'}(\mathbf{k}+\mathbf{q}) c_{a,\sigma;m'}(\mathbf{k}+\mathbf{q}) \rangle V(\mathbf{q}) \lambda_{a;m'n}(\mathbf{k},\mathbf{q}) \lambda_{a;mm'}(\mathbf{k}+\mathbf{q},-\mathbf{q})$$

We solve these equations by iterating from zero initial values. Then we add the self-energies to calculate the new Chern number. The result is summarized in Fig.4c. When the dielectric constant is large, the Chern number is 3, the same as the non-interacting case. For a fixed displacement field, increasing the interaction strength (decreasing the dielectric constant) can reduce the Chern number to 2 through a topological transition. For a large parameter region, $C = 2$ is indeed expected, consistent with the current experiment.

Finally let us discuss the nature of the observed Chern insulator. At 1/4 filling, there is one particle per moiré unit cell. Within the Hartree-Fock theory[6], the most natural ground state is a spin and valley polarized Chern insulator with $C = 2$, which is consistent with the transport measurement.

Because the two valleys are degenerate at zero magnetic field, domains can exist and cause the non-quantization of the Hall conductivity due to the chiral edge mode in the domain boundary. Upon increasing the magnetic field, the valley Zeeman coupling can align the domains and lead to perfect quantization. In the current experiment, the quantization is achieved at only 0.2 Tesla, which suggests that the valley Zeeman coupling is large, consistent with previous theoretical calculation[17].

The anomalous Hall effect in Fig.3 strongly suggests that the valley is polarized. Although the simplest ansatz within mean field theory also requires spin polarization, the present transport measurement cannot rule out more exotic Chern insulator phases with the spins in a disordered (for example, "spin liquid") or anti-ferromagnetic phase. Even for simple spin polarized scenario, non-trivial topological defects in spin space may play an important role. The skyrmion excitation carries $Q = 2e$ charge in the Chern insulator and may be the cheapest charge excitation (at small field the cheapest charge excitations may also be valley flips). In this case the activation gap is decided by the skyrmion gap. The existence of skyrmions may be reflected in a large g factor (skyrmions involve many spin flips) for response of activation gap to magnetic field. We leave it to future experiments to probe these possible interesting physics associated with the spin texture.

**Method references**
34. Zhang, F., Sahu, B., Min, H. & MacDonald, A. H. Band structure of ABC-stacked graphene trilayers. Phys. Rev. B 82, 035409 (2010).
35. Jung, J., Raoux, A., Qiao, Z. & MacDonald, A. H. Ab initio theory of moiré superlattice bands in layered two-dimensional materials. Phys. Rev. B 89, 205414 (2014).

**Data availability statement**
The data that support the findings of this study are available from the corresponding authors upon reasonable request.

**Extended data figure legends**
**Extended Data Figure 1. Magneto-transport of the Chern insulator state at $T = 1.5$ K. a,b**, Color plot of $\rho_{xx}$ and $\rho_{yx}$ as a function of carrier density and magnetic field at $D = -0.5$ V/nm and $T = 1.5$ K. The $\nu = 2$ Chern insulator state is well resolved at 1.5 K, which features a minimum of $\rho_{xx}$ and a quantized $\rho_{yx}$ emerges from 1/4 filling. **c,d**, Horizontal line cuts of **a** and **b**, respectively. $\rho_{yx}$ shows quantized Hall resistance at finite magnetic field.

**Extended Data Figure 2. Illustration of the ABC-TLG/hBN system.** The bottom hBN layer is nearly aligned with the graphene layers while the one on top is not aligned. A and B refer to the two sublattices in each of the graphene layers.

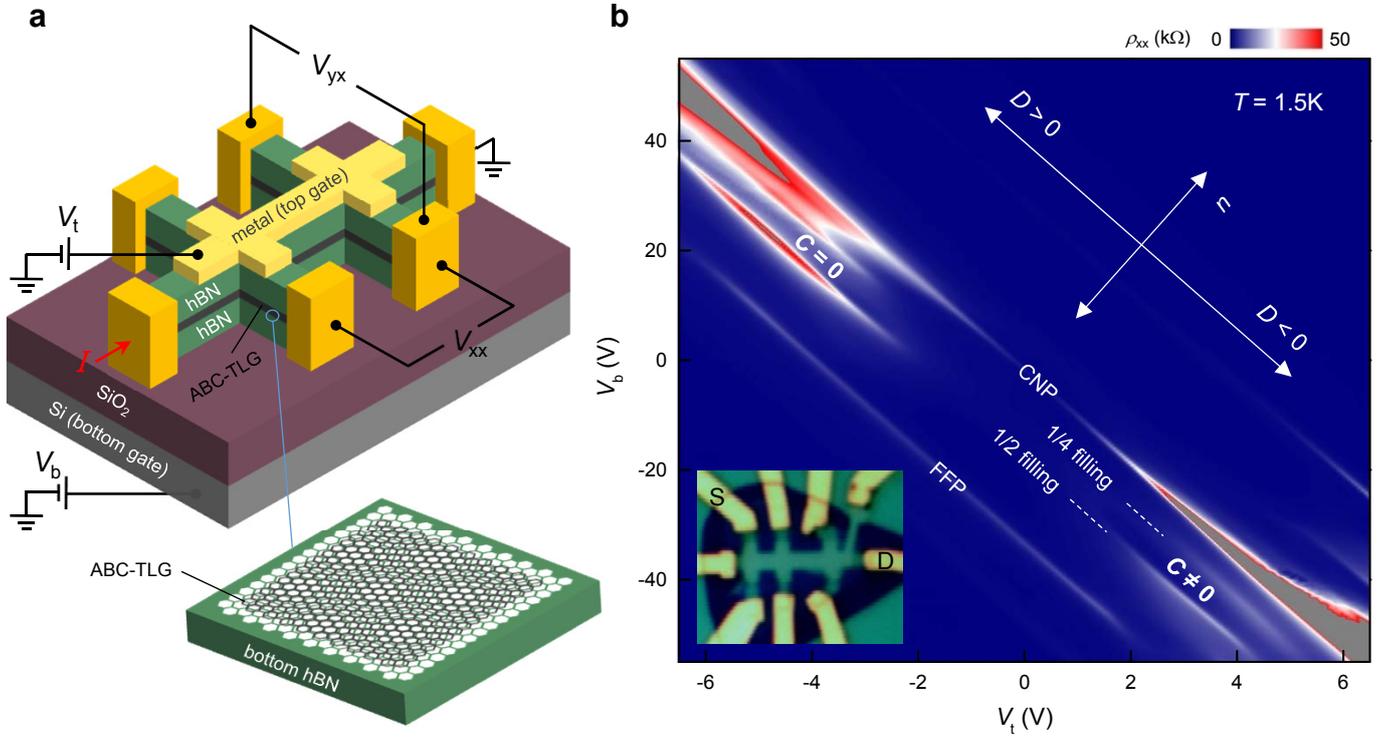

**Figure 1. TLG/hBN moiré superlattice and tunable Chern bands. a**, Schematic of the dual-gated ABC-TLG/hBN moiré superlattice Hall bar device and measurement configuration. The moiré pattern exists between TLG and bottom hBN. **b**, Color plot of the longitudinal resistivity $\rho_{xx}$ as a function of $V_t$ and $V_b$ at $T$ = 1.5 K. The arrows show the direction of changing doping ($n$) and displacement field ($D$), respectively. In addition to the band insulating states (characterized by the resistance peaks) at charge neutral point (CNP) and fully filled point (FFP), tunable correlated insulator states also emerge at 1/4 filling and 1/2 filling of the hole minibands at large displacement field $|D|$. It was predicted theoretically that the hole miniband is topological (i.e. Chern number $C \neq 0$) for $D < 0$ and trivial ($C = 0$) for $D > 0$.

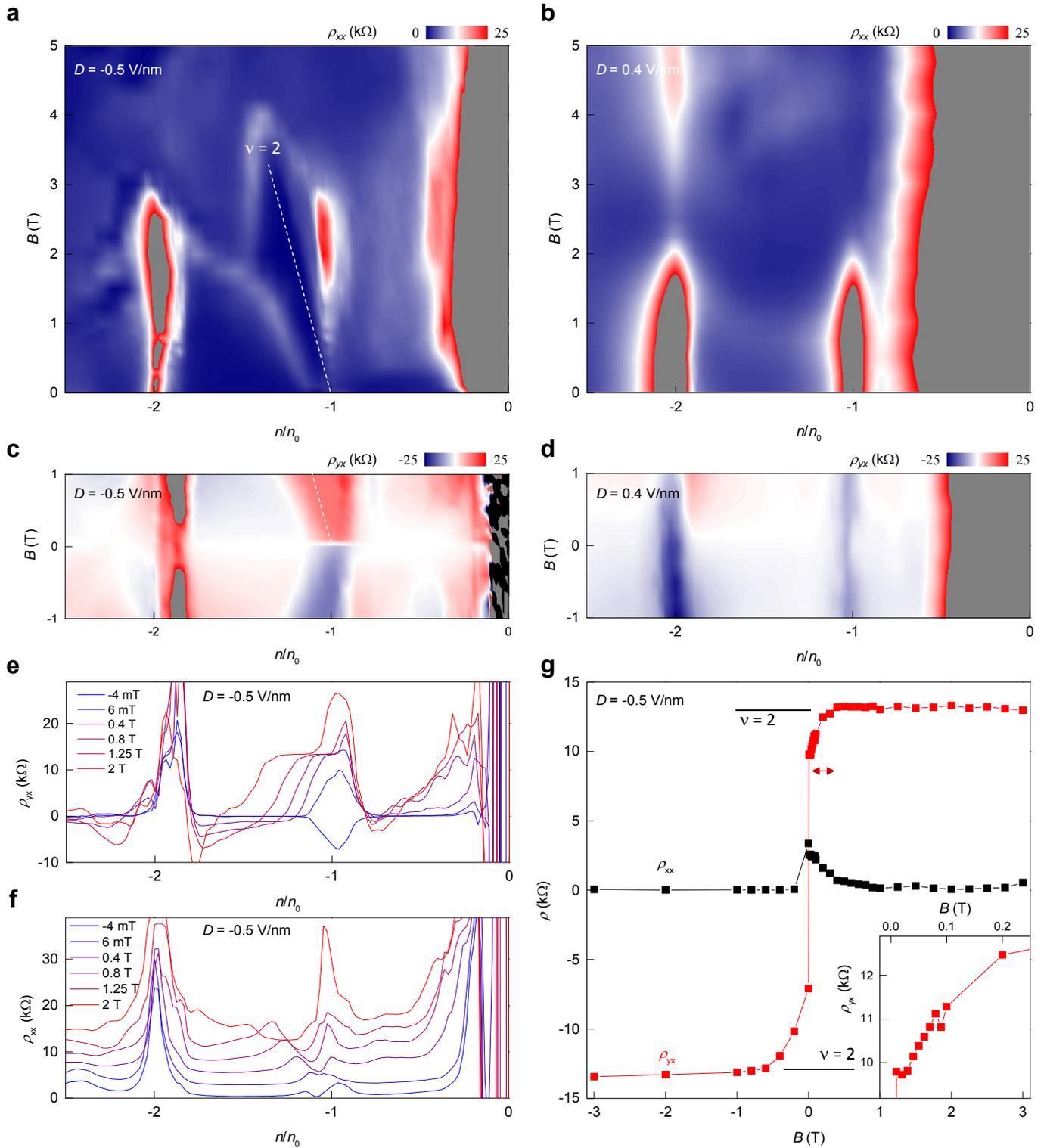

**Figure 2. Quantum Hall effect from the correlated $C = 2$ Chern insulator. a,c** Color plot of $\rho_{xx}$ and $\rho_{yx}$ as a function of carrier density and magnetic field for the topological hole miniband at $D = -0.5$ V/nm and $T = 0.06$ K. The experimental data at $T = 1.5$ K are qualitatively similar. **b,d**, Corresponding $\rho_{xx}$ and $\rho_{yx}$ plots for the trivial hole miniband at $D = 0.4$ V/nm and $T = 1.5$ K. $n_0$ corresponds to the carrier density of 1/4 filling of first miniband. No QH signature is present in the trivial hole miniband, while a $\nu = 2$ QH effect characterized by a minimum of $\rho_{xx}$ and a quantized $\rho_{yx}$ emerges from 1/4 filling of the topological hole miniband. **e,f**, Horizontal line cuts of **a** and **c**, respectively. **e** shows that $\rho_{yx}$ is well quantized beyond $B = 0.4$ T. An offset of 2.5 kΩ is applied for each stack in **f**. **g**, Line cut of $\rho_{xx}$ and $\rho_{yx}$ along the QH state (denoted by the dashed line in **a** and **e**) shows that $\rho_{yx}$ reaches a quantized value of $\nu = 2$ at 0.4 T, and a large $\rho_{yx}$ persists down to zero field. It represents a QAH state for the $C = 2$ correlated Chern insulator at 1/4 filling.

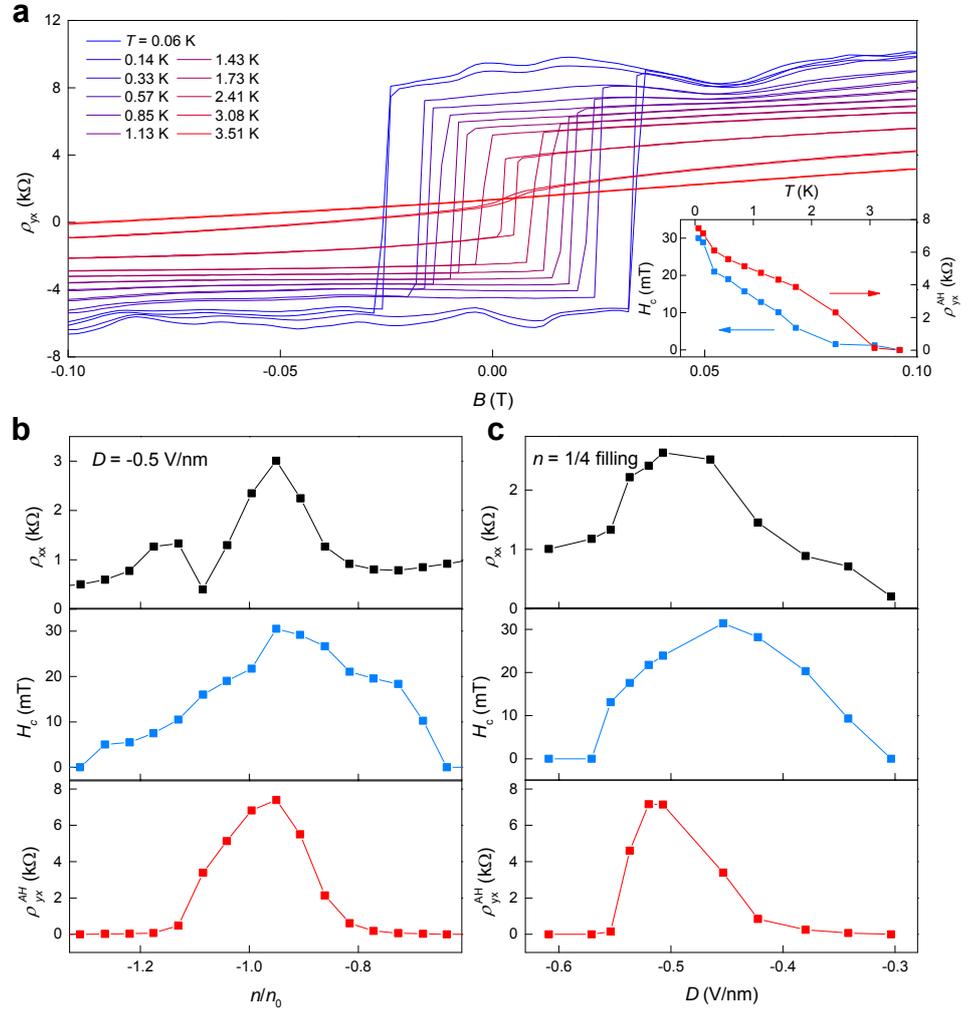

**Figure 3. Anomalous Hall effect and ferromagnetism. a**, Magnetic field dependent $\rho_{yx}$ at 1/4 filling and $D$ = -0.5 V/nm at different temperatures. The Hall resistivity displays a clear AH signal with strong ferromagnetic hysteresis. At the base temperature of $T$ = 0.06 K, the AH signal can be as high as $\rho_{yx}^{AH}$ = 8 kΩ and the coercive field is $H_c$ = 30 mT. Inset: Extracted coercive field $H_c$ and AH signal $\rho_{yx}^{AH}$ as a function of temperature. **b**, The evolution of $\rho_{xx}$, $H_c$ and $\rho_{yx}^{AH}$ as a function of hole doping at $D$ = -0.5 V/nm, $T$ = 0.06 K. The strongest AH signal is observed close to $n = n_0$. **c**, The evolution of $\rho_{xx}$, $H_c$ and $\rho_{yx}^{AH}$ as a function of the displacement field $D$ at $n = n_0$, $T$ = 0.06 K. The strongest AH signal is observed when the device is most insulating (i.e. largest $\rho_{xx}$).

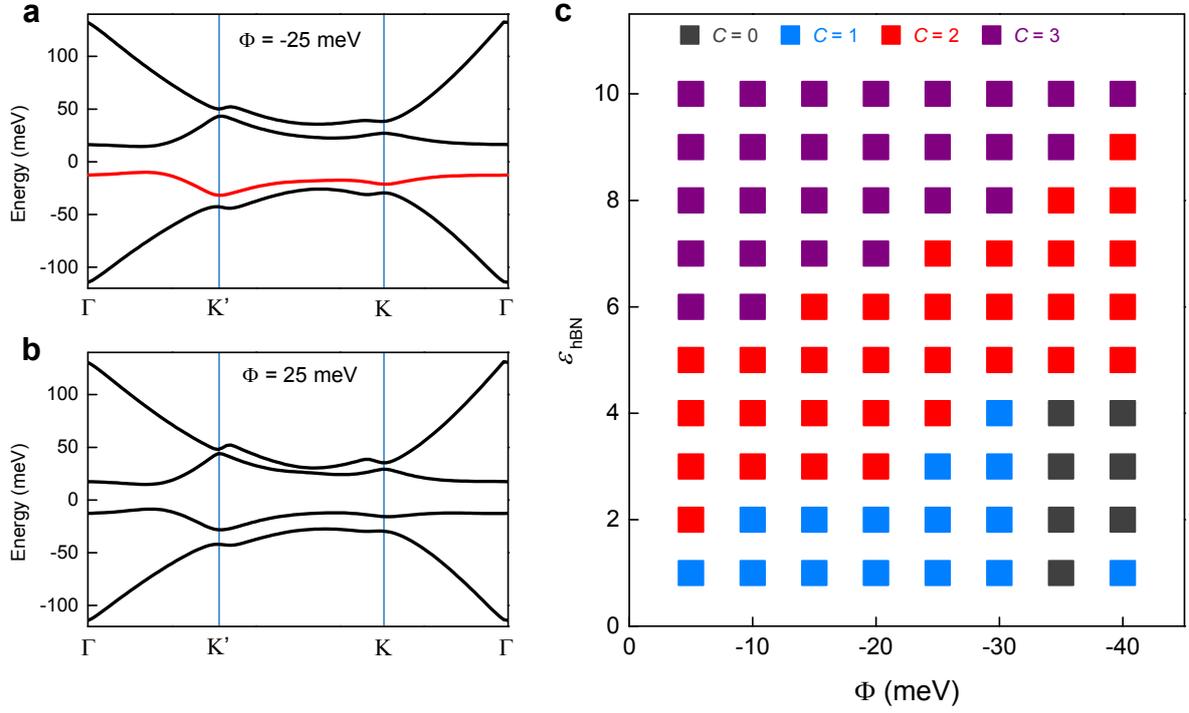

**Figure 4. Calculated Chern number including the electron-electron interaction effects. a,b**, Calculated single-particle band structure of the TLG/hBN moiré superlattice for $\Phi = -25$ meV and 25 meV, respectively. Here $\Phi$ is the energy difference between the top and bottom layer of TLG, and $\Phi = -25$ meV corresponds to vertical displacement field around $D = -0.5$ V/nm. The red line highlights the topological hole miniband for $\Phi = -25$ meV. **c**, Calculated Chern number of the hole miniband as a function of the energy difference $\Phi$ and the effective dielectric constant $\varepsilon_{hBN}$ after including the electron-electron interaction effects using the Hartree-Fock approximation. The resulting band Chern number can be 2 for parameters close to the experimental device where $\Phi \sim -25$ meV and $\varepsilon_{hBN} \sim 4$.

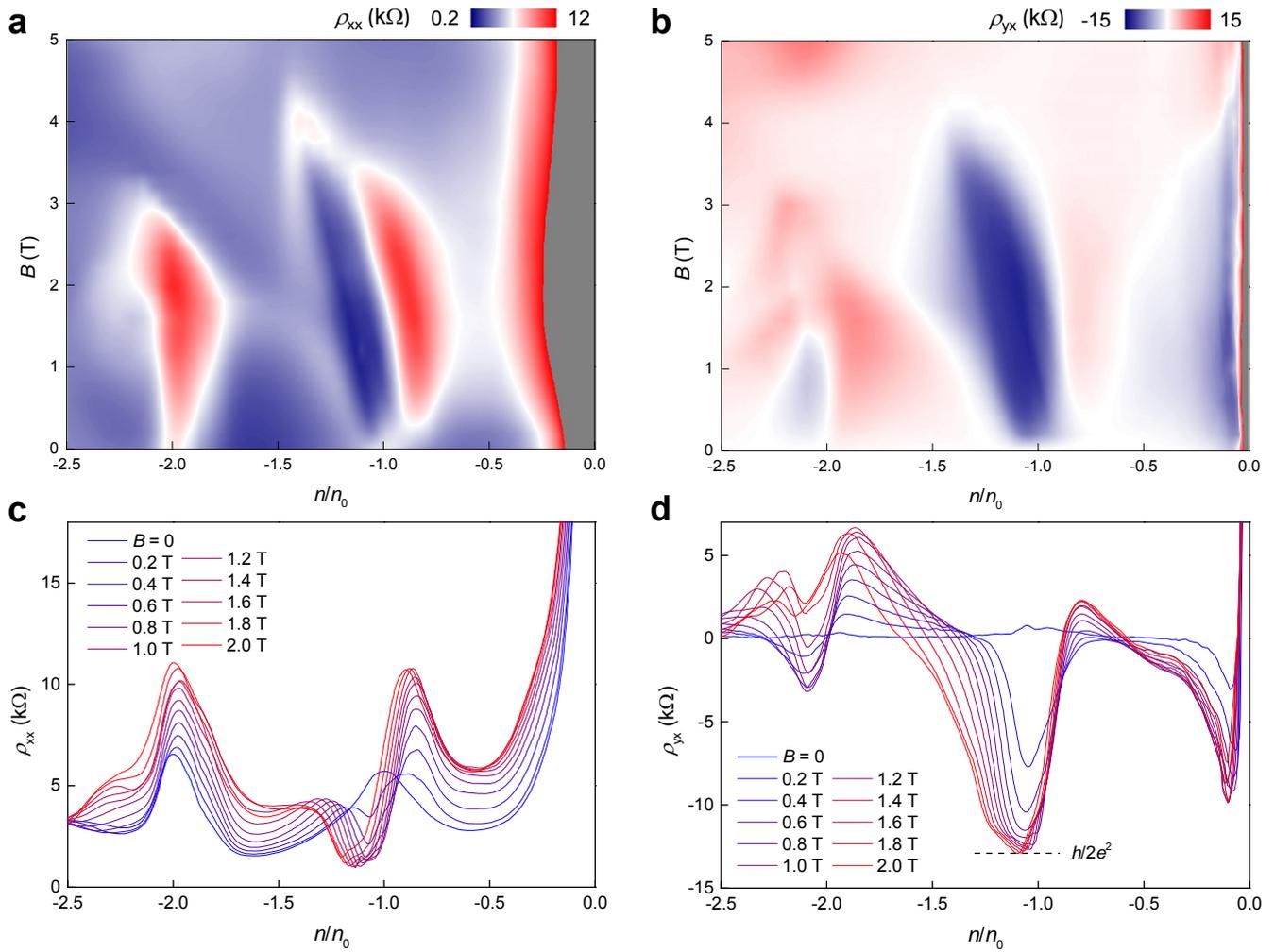

**Extended Data Figure 1. Magneto-transport of the Chern insulator state at $T$ = 1.5 K. a,b,** Color plot of $\rho_{xx}$ and $\rho_{yx}$ as a function of carrier density and magnetic field at $D$ = -0.5 V/nm and $T$ = 1.5 K. The $\nu$ = 2 Chern insulator state is well resolved at 1.5 K, which features a minimum of $\rho_{xx}$ and a quantized $\rho_{yx}$ emerges from 1/4 filling. **c,d,** Horizontal line cuts of **a** and **b**, respectively. $\rho_{yx}$ shows quantized Hall resistance at finite magnetic field.

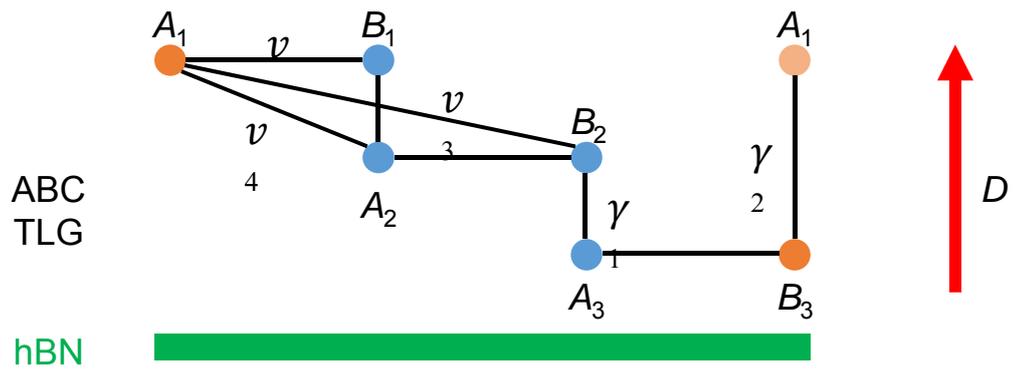

**Extended Data Figure 2. Illustration of the ABC-TLG/hBN system.** The bottom hBN layer is nearly aligned with the graphene layers while the one on top is not aligned. A and B refer to the two sublattices in each of the graphene layers.